\documentclass[prb,letterpaper,aps,floatfix,twocolumn]{revtex4-1}
\usepackage{graphicx}
\usepackage{amsmath}
\usepackage[small,bf]{subfigure}
\newcommand{\beq}{\begin{equation}}
\newcommand{\eeq}{\end{equation}}

\newcommand{\beqa}{\begin{eqnarray}}
\newcommand{\eeqa}{\end{eqnarray}}
\newcommand{\pdg}{{\vphantom \dag}}
\newcommand{\dg}{{\dag}}

\newcommand{\upa}{\uparrow}
\newcommand{\da}{\downarrow} 
\newcommand{\ra}{\rightarrow}

\begin{document}
\title{Dirac fermion duality and the parity anomaly}
\author{A.A. Burkov}
\affiliation{Department of Physics and Astronomy, University of Waterloo, Waterloo, Ontario 
N2L 3G1, Canada} 
\date{\today}
\begin{abstract}
We present a derivation of the recently discovered duality between the free massless $(2+1)$-dimensional Dirac fermion and 
QED$_3$. Our derivation is based on a regularized lattice model of the Dirac fermion and is similar to the more familiar derivation of the boson-vortex duality. It also highlights the important role played by the parity anomaly, which is somewhat less obvious in other discussions of this duality in the literature.  
\end{abstract}
\maketitle
\section{Introduction}
\label{sec:1}
The concept of duality plays an important role in both classical and quantum many-body physics. 
Duality typically maps a strongly-coupled phase of one theory onto a weakly-coupled phase of the dual theory,
thus relating seemingly very different models. 
A particularly prominent example of duality is the boson-vortex duality.~\cite{Peskin_duality,Kogut_RMP}
Its classic applications include theory of the Kosterlitz-Thouless transition in the two-dimensional (2D) XY 
model,~\cite{Berezinski,KT,Kogut_RMP} demonstration of the inverted-XY nature of the superconductor-normal transition in 
type-II superconductors,~\cite{Dasgupta-Halperin} superconductor-insulator transition in thin films,~\cite{Fisher-Lee,Fisher_SI,Galitski05} and the quantum Hall effect.~\cite{Lee-Fisher,Lee-Kane,ZHK}
It has also provided important insights into quantum phase transitions, which can not be described by the Landau-Ginzburg-Wilson theory (deconfined criticality).~\cite{DQCP,Burkov_DQCP,Burkov05}
Recently, boson-vortex duality has been extended to fermion-fermion,~\cite{Wang-Senthil,Metlitski-Vishwanath,Alicea16}
and boson-fermion~\cite{Seiberg16,Karch16,Raghu18} cases. 
This has already led to significant new developments in the theory of the half-filled Landau level~\cite{Son15,Geraedts16} and deconfined criticality.~\cite{Senthil17,He18}

In this paper we provide a derivation of the duality between free massless $(2+1)$-dimensional Dirac fermion and massless 
QED$_3$,~\cite{Wang-Senthil,Metlitski-Vishwanath,Alicea16} which is different from what has been presented in the literature so far. 
Our derivation is closely analogous to the much more familiar derivation of the boson-vortex duality, which might have some pedagogical value. 
It also highlights the crucial 
role of the parity anomaly, which is somewhat less transparent in the existing discussions of this duality in the literature. 

The statement of the free Dirac fermion to QED$_3$ duality refers to equivalence, at long distances and low energies, of the following euclidean field theories
\beq
\label{eq:1}
{\cal L} = \bar \psi \gamma^{\mu}(\partial_{\mu} - i A_{\mu}) \psi, 
\eeq
and 
\beq
\label{eq:2}
{\cal L}_d = \bar \psi_d \gamma^{\mu} (\partial_{\mu} - i a_{\mu})\psi_d + \frac{i}{4 \pi} \epsilon^{\mu \nu \lambda} A_{\mu} \partial_{\nu} a_{\lambda}. 
\eeq
Here $\bar \psi = \psi^\dg \gamma^0$ is the Dirac adjoint, $\gamma^{\mu}$ are hermitian gamma matrices (which are Pauli matrices in this case), $A_{\mu}$ is the source electromagnetic gauge potential, $a_{\mu}$ is a dynamical gauge field, whose curl is proportional to the free Dirac fermion 3-current, and $\hbar = c = e =1$ units are used here 
and throughout this paper. 

The duality between Eqs.~\eqref{eq:1} and \eqref{eq:2} is easy to establish if mass terms are added to both Lagrangians, in which case Eq.~\eqref{eq:1} describes the quantum Hall plateau transition at which $\sigma_{xy}$ jumps by $1/ 2 \pi$.~\cite{Haldane88,Ludwig94}
Integrating out the free Dirac fermion with mass $m$ in Eq.~\eqref{eq:1} produces a half-quantized Chern-Simons term for the electromagnetic field
\beq
\label{eq:3}
{\cal L} = - i \frac{\textrm{sign}(m)}{8 \pi} \epsilon^{\mu \nu \lambda} A_{\mu} \partial_\nu A_{\lambda} + \ldots,
\eeq
where $\ldots$ denote the Maxwell part of the response $\propto F_{\mu \nu} F^{\mu \nu}$, which is subdominant to the Chern-Simons 
term at long distances and long times. 
Analogously, integrating out dual fermions $\psi_d$  with mass $m_d$ in Eq.~\eqref{eq:2} gives 
\beq
\label{eq:4}
{\cal L}_d = - i \frac{\textrm{sign}(m_d)}{8 \pi} \epsilon^{\mu \nu \lambda} a_{\mu} \partial_\nu a_{\lambda} + \frac{i}{4 \pi} \epsilon^{\mu \nu \lambda} A_{\mu} \partial_{\nu} a_{\lambda} + \ldots, 
\eeq
where $\ldots$ again refer to the subdominant Maxwell term for the gauge field $a_{\mu}$. 
Further integrating out $a_{\mu}$ gives 
\beq
\label{eq:5}
{\cal L}_d =  i \frac{\textrm{sign}(m_d)}{8 \pi} \epsilon^{\mu \nu \lambda} A_{\mu} \partial_\nu A_{\lambda} + \ldots,
\eeq
Thus ${\cal L}$ and ${\cal L}_d$ are equivalent in the long wavelength and low energy limit if the masses $m$ and $m_d$ 
are taken to have opposite signs. 
As seen from the above arguments, this statement is a straightforward (at least in the large mass limit) consequence of the well-known self-duality of the Chern-Simons term~\cite{Zee_book} and the fact that it dominates all other responses at long distances and long times, which leads to the universal quantum Hall insulator response of a massive $(2+1)$-dimensional Dirac fermion. 
What is less trivial, however, is that this duality continues to hold in the massless limit of Eqs.~\eqref{eq:1} and \eqref{eq:2}, 
in particular when Eq.~\eqref{eq:1} describes the surface state of a three-dimensional (3D) time-reversal (TR)-invariant 
topological insulator (TI). 
This has been established originally in Refs.~\onlinecite{Wang-Senthil,Metlitski-Vishwanath,Alicea16} and this is what we are 
going to rederive in this paper. 
Our derivation is somewhat similar in spirit to the recent derivation of the boson-fermion duality in Ref.~\onlinecite{Raghu18}, in the sense 
that we also use a lattice-regularized euclidean theory, but many of the details are different. 

The rest of the paper is organized as follows. 
In Section~\ref{sec:2} we derive the Dirac fermion duality using the slave-rotor gauge theory.~\cite{Georges_SR}
We conclude in Section~\ref{sec:3} with a discussion of our results. 

\section{Derivation of the duality}
\label{sec:2}
The low-energy Lagrangians of Eqs.~\eqref{eq:1} and \eqref{eq:2} are not sharply defined due to ultraviolet divergences and the parity anomaly discussed below.
We will instead start from a well-defined lattice Hamiltonian, which at low energies reduces to Eq.~\eqref{eq:1}. 
We will take this Hamiltonian to have the following form in momentum space:
\beq
\label{eq:6}
H = \sum_k c^\dg_k \sigma^\mu d_{\mu}(k) c^\pdg_k, 
\eeq
where $\sigma$ are Pauli matrices, $c^\dg_k$ creates an electron with crystal momentum $k$ (spin indices have been made implicit)
and 
\beqa
\label{eq:7}
&&d_x(k) = \sin k_x, \,\, d_y(k) = \sin k_y, \nonumber \\
&&d_z(k) = m_0 + m_1 (2 - \cos k_x - \cos k_y). 
\eeqa
To make the notation more economical we will use the same Greek indices $\mu, \nu, \lambda$ for both purely spatial directions, as in Eq.~\eqref{eq:6}, and later also for the temporal direction as well, the range of implicit summation 
in every case will be clear from the context. 

When the parameter $m_0 = 0$ while $m_1 \neq 0$, Eq.~\eqref{eq:6} describes a single massless Dirac fermion at the $\Gamma$-point in the Brillouin zone (BZ). 
Note that while the low-energy Hamiltonian with $m_0 = 0$
\beq
\label{eq:8}
H = \sum_k c^\dg_k \sigma^\mu k_\mu c^\pdg_k,
\eeq
possesses both parity and TR symmetries, regularized Hamiltonian of a single massless 2D Dirac fermion necessarily breaks both, since the parameter $m_1$ must be nonzero in order to avoid extra massless fermions at the edge of the BZ. 
In other words, it is impossible to put a single (or, generally, any odd number) 2D Dirac fermion on a lattice without violating 
both parity and TR symmetry, even though both seem to be present to linear order in momentum. This property of 2D Dirac fermions is known as the parity anomaly~\cite{Semenoff84,Redlich84,Haldane88} and it will play an important role in what follows. 

The electronic structure topology of the Dirac Hamiltonian Eq.~\eqref{eq:6} may be characterized by the Chern number of the valence 
band with the band energy $\epsilon(k) = - d(k) \equiv - \sqrt{d^{\mu} d_{\mu}}$.  It is easily found by evaluating the total circulation of the Berry connection vector 
\beq
\label{eq:9}
{\cal A}_{\mu}(k) = \frac{d_x \partial_{\mu} d_y - d_y \partial_{\mu} d_x}{2 d (d - d_z)}, 
\eeq
around the points in the BZ, at which ${\cal A}_{\mu}(k)$ is singular. 
These points are located either at the BZ center $k = (0,0)$ or at the edges $k = (0,\pi), (\pi, 0), (\pi, \pi)$, and arise either as a result 
of band touchings, or in a fully gapped bandstructure, when bands carry nonzero Chern numbers. In the latter case, the singular points may be thought of as points, where a Dirac string enters or leaves the BZ. 

After a straightforward calculation, one obtains the following result for the Chern number
\beqa
\label{eq:10}
&&C = - \frac{1}{2} \left[1 + \textrm{sign}(m_0) \right] \nonumber \\
&+&\frac{1}{2} \left[1 + 2\textrm{sign}(m_0 + 2 m_1) - \textrm{sign}(m_0 + 4 m_1) \right]. 
\eeqa
The first term in Eq.~\eqref{eq:10} arises from the circulation of the Berry connection around the $\Gamma$-point, while the rest 
comes from singular points at the edge of the BZ. 
When $m_0 = 0$, which is the case of interest to us, we have $C = \textrm{sign}(m_1)/2$, i.e. the Chern number is half-integer, 
which is a consequence of the singularity of the electronic structure due to the band touching at the $\Gamma$-point. 

After an inverse lattice Fourier transform 
\beq
\label{eq:11}
c^\dg_k = \frac{1}{\sqrt{N}} \sum_r e^{i k \cdot r} c^\dg_r, 
\eeq
and coupling to an external gauge potential $A_{r \mu}$, defined on the lattice links $(r \mu)$, Eq.~\eqref{eq:6} becomes
\beqa
\label{eq:12}
H&=&\sum_r \left[- \frac{i}{2} c^\dg_r (\sigma^\mu - i m_1 \sigma^z) c_{r + \mu} e^{- i A_{r \mu}}  + h.c. \right. \nonumber \\
&+&\left.(m_0 + 2 m_1) c^\dg_r \sigma^z c^\pdg_r - i A_{r 0} c^\dg_r c_r \right]. 
\eeqa
Integrating out fermions in $H$ induces a Chern-Simons term in the electromagnetic field Lagrangian
\beq
\label{eq:13}
{\cal L} = i \frac{C}{4 \pi} \epsilon^{\mu \nu \lambda} A_{\mu} \partial_{\nu} A_{\lambda} + \ldots,
\eeq
where $\ldots$ represent other, ``nontopological" terms, including nonlocal response, generated by the gapless fermions. 
In the massless $m_0 = 0$ limit, and ignoring the nontopological terms, Eq.~\eqref{eq:13} becomes
\beq
\label{eq:14}
{\cal L}_{CS}(A) = i \frac{\textrm{sign}(m_1)}{8 \pi} \epsilon^{\mu \nu \lambda} A_{\mu} \partial_{\nu} A_{\lambda}. 
\eeq
This half-quantized Chern-Simons term, generated by a massless regularized 2D Dirac fermion, is a direct manifestation of the 
parity anomaly. 

If we take the low-energy Dirac fermion in Eq.~\eqref{eq:1} 
to represent a TR-invariant phase, which can exist as a surface state of a 3D TI, we must subtract exactly the same half-quantized Chern-Simons term from the regularized theory in order to cancel ${\cal L}_{CS}(A)$, since it can not be present in a TR-invariant system.~\cite{Redlich84}
Physically this subtracted term may be thought of as being generated by the bulk in a semi-infinite sample.~\cite{Mulligan13}
Another, more explicit and transparent solution, is to consider a TI sample in the form of a slab with two (top and bottom) surfaces,
which are not mixed at the $\Gamma$-point, but get hybridized away from it, which models the effect of merging 
of the surface states with the bulk at finite momenta.~\cite{Franz12}
The Hamiltonian of this system may be written as
\beq
\label{eq:14.1}
H = \sum_k c^\dg_k [\tau^z (\sigma^x \sin k_x + \sigma^y \sin k_y) + \tau^x \Delta(k)] c^\pdg_k, 
\eeq
where the eigenvalues of $\tau^z$ represent the top (T) and bottom (B) surfaces of the TI film and
\beq
\label{eq:14.2}
\Delta(k) = m_1 (2 - \cos k_x - \cos k_y). 
\eeq
A unitary transformation 
\beqa
\label{eq:14.3}
&&c^\dg_{k \upa T} \ra c^\dg_{k \upa T},\,\, c^\dg_{k \da T} \ra c^\dg_{k \da T}, \nonumber \\
&&c^\dg_{k \upa B} \ra c^\dg_{k \upa B},\,\, c^\dg_{k \da B} \ra - c^\dg_{k \da B}, 
\eeqa
brings the Hamiltonian to the form 
\beq
\label{eq:14.4}
H = \sum_k c^\dg_k [\sigma^x \sin k_x + \sigma^y \sin k_y + \sigma^z \tau^x \Delta(k)] c^\pdg_k.
\eeq
This describes two independent lattice Dirac fermions, corresponding to the two eigenvalues of $\tau^x$. 
The two Dirac fermions are related to each other by time reversal and their parity anomalies cancel each other. 
We will implicitly assume henceforth that we are dealing with a system, described by Eq.~\eqref{eq:14.4}, while explicitly working with a single 
lattice Dirac fermion. 

In order to find the dual of Eq.~\eqref{eq:12} we will start by representing the fermion creation operator as a product of a 
charged boson creation operator and a neutral fermion creation operator as follows~\cite{Georges_SR}
\beq
\label{eq:15}
c^\dg_r = e^{i \theta_r} f^\dg_r. 
\eeq
Here $e^{i \theta_r}$ is the creation operator of a spinless boson, carrying the original electron's charge, while $f^\dg_r$ creates 
a neutral fermion, which carries the spin. 
Conjugate to the phase $\theta_r$ is the boson number operator $n_r$, with 
\beq
\label{eq:16}
[\theta_r, n_r] = i.
\eeq 
We will take $n_r = -1$ to represent the state with no electrons on site $r$, $n_r = 0$ the two states with one electron and 
$n_r = 1$ the state with two electrons at $r$. 
This means that the parton operators must satisfy the following constraint
\beq
\label{eq:17}
f^\dg_r f^\pdg_r  = n_r +1. 
\eeq
In the parton language a free Dirac fermion corresponds to the superfluid phase of the charged bosons with 
$\langle e^{i \theta_r} \rangle \neq 0$, which implies $c^\dg_r \sim f^\dg_r$. 
The phase with $\langle e^{i \theta} \rangle = 0$ would correspond to a Mott insulator, which we will not discuss here. 

This representation of the fermion operators leads to the following imaginary time action
\beqa
\label{eq:18}
&&S = \int_0^{\beta} d \tau \sum_r\left[f^\dg_r \partial_{\tau} f^\pdg_r - i n_r (\partial_{\tau} \theta_r + A_{0 \tau}) \right. \nonumber \\
&+&\left. i \lambda_r (f^\dg_r f^\pdg_r - n_r -1) + 2 m_1 f^\dg_r \sigma^z f^\pdg_r \right. \nonumber \\
&-&\left. \frac{i}{2} f^\dg_r (\sigma^\mu - i m_1 \sigma^z) f_{r + \mu} e^{- (\Delta_{\mu} \theta_r + A_{r \mu})}  + h.c. \right], 
\eeqa
where $\lambda_r$ is a Lagrange multiplier field, enforcing the constraint \eqref{eq:17} and $\Delta_{\mu} \theta_r \equiv
\theta_{r+\mu} - \theta_r$. 
We now decouple the last term in the action, which couples the bosons and the fermions, by a Hubbard-Stratonovich transformation and obtain~\cite{Lee_SR,Paramekanti09}
\beqa
\label{eq:19}
S&=&\int_0^{\beta} d \tau \sum_r\left[f^\dg_r (\partial_{\tau} + i b_{r 0}) f^\pdg_r - i n_r (\partial_{\tau} \theta_r + A_{r 0} + b_{r 0})  \right.\nonumber \\
&-&\left.\frac{i \chi}{2} f^\dg_r (\sigma^\mu - i m_1 \sigma^z) f_{r + \mu} e^{i b_{r \mu}} + h.c.\right. \nonumber \\
&-&\left.\chi \cos(\Delta_{\mu} \theta_r + A_{r \mu} + b_{r \mu}) + 2 m_1 f^\dg_r \sigma^z f^\pdg_r \right], 
\eeqa
where the gauge field $b_{r \mu}$ is the phase of the Hubbard-Stratonovich field defined on the lattice link $(r \mu)$, 
while its magnitude $\chi$ has been taken to be constant (equal to its saddle-point value).
We have also identified the Lagrange multiplier field $\lambda_r$ with the temporal component of the gauge field $b_{\mu}$. 

We now use the standard Villain approximation to the boson kinetic energy term in Eq.~\eqref{eq:19}.~\cite{Villain_duality} 
Namely, we have
\beq
\label{eq:20}
e^{\chi \cos(\Delta_{\mu} \theta_r + A_{r \mu} + b_{r \mu})} \sim \sum_{J_{r \mu}} e^{- i J_{r \mu} (\Delta_{\mu} \theta_r + A_{r \mu} + 
b_{r \mu}) - \frac{1}{2 \chi} J_{r \mu}^2}, 
\eeq
where $J_{r \mu}$ are integer variables, defined on the lattice links, which represent components of the boson current. 
The imaginary time action becomes
\beq
\label{eq:21}
S = S_f(b) + \sum_{r \tau} \left[i J_{r \mu}(\Delta_{\mu} \theta_r + A_{r \mu} + b_{r \mu}) + \frac{1}{2 \chi} J_{r \mu}^2 \right],
\eeq
where 
\beqa
\label{eq:21.5}
S_f(b)&=&\sum_{r \tau}\left[f^\dg_r (\Delta_\tau + i b_{r 0}) f^\pdg_r + 2 m_1 f^\dg_r \sigma^z f^\pdg_r\right. \nonumber \\
&-&\left.\frac{i \chi}{2} f^\dg_r (\sigma^\mu - i m_1 \sigma^z) f_{r + \mu} e^{i b_{r \mu}} + h.c.\right]. 
\eeqa
Here we have switched from integration over imaginary time to summation, set the time step to unity for simplicity
and identified the boson charge $n_r$ with the temporal component of the boson current $J_{r0}$. 
We have also added a term $J_{r 0}^2/2 \chi$, which was not there in Eq.~\eqref{eq:19}. 
The justification for this is that such a term will be generated by integrating out high-energy modes in \eqref{eq:19}.
This coarse-graining procedure will be explicitly carried out when we go back from the regularized lattice theory to low-energy continuum theory at the end. 

Integrating out the boson phase variable in Eq.~\eqref{eq:21}, we obtain the boson charge conservation law
\beq
\label{eq:22}
\Delta_{\mu} J_{r \mu} = 0,
\eeq
which may be solved as 
\beq
\label{eq:23}
J_{\mu} = \epsilon^{\mu \nu \lambda} \Delta_{\nu} a_{\lambda}, 
\eeq
where $a_{\lambda}$ is an integer-valued gauge field, defined on the links of the dual lattice. From now on we will suppress the spatial 
and temporal indices to avoid multiple indices referring to the direct and the dual lattices. 

At this point we would like to integrate out the gauge field $b_{\mu}$. This is not straightforward, since it is coupled to gapless fermions $f$, 
which may also not be integrated out due to the gaplessness (more precisely, they may be integrated out but this will generate nonlocal terms). 
To deal with this problem we will use the following process. 
We first separate the fermion modes into low- and high-energy ones: the low-energy modes arise from a small neighborhood of the 
Dirac point with $k < \Lambda \ll 1$ and the high-energy modes correspond to the rest of the Brillouin zone. 
The high-energy modes may now be integrated out straightforwardly. This produces a half-quantized Chern-Simons term for the gauge field $b_{\mu}$, 
plus Maxwell-type terms that may be ignored
\beq
\label{eq:24}
S = \tilde S_f(b) + \sum \left[{\cal L}_{CS}(b) + i \epsilon^{\mu \nu \lambda} (A_{\mu} + b_{\mu}) \Delta_{\nu} a_{\lambda} + \ldots \right],
\eeq
where $\tilde S_f(b)$ is the long-wavelength limit of $S_f(b)$, given by
\beq
\label{eq:24.5}
\tilde S_f(b) = \int d^3 x \bar f \gamma^{\mu} (\partial_{\mu} + i b_{\mu}) f. 
\eeq

The next step is to use the well-known fact that the Chern-Simons term acts as a mass term for the gauge field.~\cite{Deser82}
We first rewrite the action as
\beqa
\label{eq:25}
S&=&\tilde S_f(b) + \sum \left[{\cal L}_{CS}(b + 4 \pi a) - 2 \pi i \epsilon^{\mu \nu \lambda} a_{\mu} \Delta_{\nu} a_{\lambda}  \right. \nonumber \\
&+&\left. i \epsilon^{\mu \nu \lambda} A_{\mu} \Delta_{\nu} a_{\lambda} + \ldots \right]. 
\eeqa
Since ${\cal L}_{CS}(b + 4 \pi a)$ acts as a mass term, we may now use the saddle-point approximation to integrate out $b_{\mu}$, 
which simply amounts to replacing $b_{\mu} \rightarrow - 4 \pi a_{\mu}$, assuming that the correction to the saddle point due to $\tilde S_f(b)$ 
may be ignored (this correction will only generate interaction terms between the fermions, but these will be generated anyway due to the coupling 
of the fermions to the fluctuating gauge field $a_{\mu}$). 
This gives 
\beqa
\label{eq:26}
&&S = \tilde S_f(-4 \pi a) \nonumber \\
&+&\sum \left[-2 \pi i \epsilon^{\mu \nu \lambda} a_{\mu} \Delta_{\nu} a_{\lambda}  + i \epsilon^{\mu \nu \lambda} A_{\mu} \Delta_{\nu} a_{\lambda} + \ldots \right]. 
\eeqa

We now relax the integer constraint on the gauge field $a_{\mu}$ by introducing a potential $- t \cos(\Delta_{\mu} \varphi + 2 \pi a_{\mu})$, where the new phase variable $\varphi$ is needed to keep the action gauge invariant.
Rescaling $4 \pi a_{\mu} \rightarrow a_{\mu}$ we finally obtain the dual imaginary time action 
\beqa
\label{eq:27}
S&=&\tilde S_f(-a) + \sum \left[- {\cal L}_{CS}(a) + \frac{i}{4 \pi} \epsilon^{\mu \nu \lambda} A_{\mu} \Delta_{\nu} a_{\lambda} \right. \nonumber \\
&-&\left.t \cos(\Delta_{\mu} \varphi + a_{\mu}/2) + \ldots \right]. 
\eeqa
The boson field $e^{i \varphi}$ is dual to $e^{i \theta}$ and thus represents vortices in the original slave-rotor field. 
Since the free Dirac fermion corresponds to the superfluid phase of $e^{i \theta}$, the vortex field is in the gapped insulator phase 
and may be integrated out. This will generate only extra subdominant Maxwell terms for the gauge field $a_{\mu}$ in the long-wavelength limit, which we will ignore. 
The third term in Eq.~\eqref{eq:27} couples two flux quanta of the field $a_{\mu}$ to every electron charge. This is also the reason 
why the dual vortex field carries a gauge charge of $1/2$, as seen from the fourth term in \eqref{eq:27}: a vortex in $e^{i\varphi}$, which is just the original electron charge, then clearly carries a flux of $4 \pi$, i.e. two flux quanta, as it should. 
At the same time the dual fermions $f$ carry a unit gauge charge, which means that each dual fermion corresponds to a double vortex in the 
original electron field $c$. 
The half-quantized Chern-Simons term ${\cal L}_{CS}(a)$ should be regarded as arising from lattice regularization of the low-energy theory of Eq.~\eqref{eq:2} and should be omitted when the continuum limit is taken. 
Taking the continuum limit of all the terms in Eq.~\eqref{eq:27} thus indeed reproduces Eq.~\eqref{eq:2}. 
This completes the derivation of the Dirac fermion duality. 
\section{Discussion and conclusions}
\label{sec:3}
Since the above derivation, while mostly straightforward and following standard steps, familiar from the boson-vortex duality, is 
also somewhat formal, and is not without some subtleties, we will now summarize a few of its important details. 

First, the parton representation \eqref{eq:15} was only necessary in order to carry out the standard duality step of representing 
the conserved charge current as a curl of a dynamical gauge field $a_{\mu}$. 
Second, perhaps the most nontrivial step of the duality transformation is the step of integrating out the gauge field $b_{\mu}$ in 
Eqs.~\eqref{eq:24}-\eqref{eq:26}. 
The difficulty here arises from the fact that $b_{\mu}$ is coupled to massless fermions. 
In order to deal with this, we divided the fermion modes into low- and high-energy parts and integrated out the high-energy modes. 
This produced the half-quantized Chern-Simons term ${\cal L}_{CS}(b)$ (along with subdominant Maxwell terms), which acted as a topological mass term for the gauge field $b_{\mu}$. 
This fact had allowed us to integrate $b_{\mu}$ out using the saddle-point approximation.

One may worry that the half-quantized Chern-Simons term ${\cal L}_{CS}(b)$, which has played a significant role in this derivation, 
is only a consequence of the lattice regularization of the Dirac fermion and is thus artificial. 
Indeed, its direct analog ${\cal L}_{CS}(A)$ is cancelled between the two time-reversed Dirac fermions in Eq.~\eqref{eq:14.4}. 
However, there is an important physical distinction between the corresponding gauge fields, $A_{\mu}$ and $b_{\mu}$. 
$A_{\mu}$ corresponds to the external electromagnetic field and is coupled equally to both Dirac fermions in Eq.~\eqref{eq:14.4}. 
In contrast, $b_{\mu}$ is an internal gauge field, whose role is to glue together the charge and the spin of each Dirac fermion independently. 
Thus the terms ${\cal L}_{CS}(b)$ are not mutually cancelled between the two Dirac fermions, since they involve two independent gauge fields, and 
their appearance is in fact an inevitable consequence of gauge invariance. 

In conclusion, we have rederived the duality between free massless $(2+1)$-dimensional Dirac fermion and QED$_3$ using the 
slave-rotor representation of the electron operators, combined with the standard duality transformation of the bosonic rotor variables. 
One advantage of this derivation, apart from its simplicity, is that the crucial role played by the topological properties of 2D Dirac fermions, namely the parity anomaly, is particularly clear. 

\begin{acknowledgments}
We acknowledge useful discussions with Yin-Chen He and Arun Paramekanti. Financial support was provided by Natural Sciences and Engineering Research Council (NSERC) of Canada. 
\end{acknowledgments}
\bibliography{references}
\end{document}